\documentclass[aps,prl,twocolumn,showpacs,superscriptaddress,groupedaddress]{revtex4}
\usepackage{blindtext}
\usepackage{amsmath,amssymb}    
\usepackage{graphicx}   
\usepackage{verbatim}   
\usepackage{color}      
\usepackage{subfigure}  
\usepackage{bm}
\usepackage{textgreek}

\preprint{\bf PREPRINT}

\begin{document}

\title{Buckling of an elastic ridge:\\
competition between wrinkles and creases }
\author{C.~Lestringant} \affiliation{Sorbonne Universit\'es,
UPMC Univ Paris 06, CNRS, UMR 7190,
Institut \textdelta 'Alembert, F-75005, Paris, France}
\author{C.~Maurini} \affiliation{Sorbonne Universit\'es,
UPMC Univ Paris 06, CNRS, UMR 7190,
Institut \textdelta 'Alembert, F-75005, Paris, France}
\author{A.~Lazarus} \affiliation{Sorbonne Universit\'es,
UPMC Univ Paris 06, CNRS, UMR 7190,
Institut \textdelta 'Alembert, F-75005, Paris, France}
\author{B.~Audoly} \affiliation{Laboratoire de M\'ecanique des Solides, CNRS,
UMR  7649, D\'epartement de M\'ecanique,
\'Ecole Polytechnique, 91128 Palaiseau CEDEX, France}

\vskip 0.25cm

\begin{abstract}
We investigate the elastic buckling of a triangular prism made of
a soft elastomer.  A face of the prism is bonded to a stiff slab
that imposes an average axial compression.
We observe two possible buckling modes which are localized along the
free ridge.  For ridge angles $\phi$ below a critical value
$\phi^\star\approx 90^\circ$ experiments reveal an extended sinusoidal
mode, while for $\phi$ above $\phi^\star$ we observe a series of
creases progressively invading the lateral faces starting from the
ridge.  A numerical linear stability analysis is set up using the
finite-element method and correctly predicts the sinusoidal mode for
$\phi \leq \phi^\star$, as well as the associated critical strain
$\epsilon_{\mathrm{c}}(\phi)$.  The experimental transition at
$\phi^\star$ is found to occur when this critical strain
$\epsilon_{\mathrm{c}}(\phi)$ attains the value
$\epsilon_{\mathrm{c}}(\phi^\star) = 0.44$ corresponding to the
threshold of the sub-critical surface creasing instability.  Previous
analyses have focused on elastic crease patterns appearing on planar
surfaces, where the role of scale-invariance has been emphasized; our
analysis of the elastic ridge provides a different perspective, and
reveals that scale-invariance is not a sufficient condition for
localization.
\end{abstract}

\maketitle

%

In extended systems, elastic instabilities generally produce smooth
patterns having a well-defined wavelength.  There are numerous
examples involving an elastic
beam~\cite{Savin-Kurpios-EtAl-On-the-growth-and-form-of-the-gut-2011}
or a thin film~\cite{Biot57,Sultan_Boudaoud, Cao_Hutch_Wrinkling} on an elastic
foundation, a bulk elastic material with inhomogeneous elastic
properties~\cite{Lee-Triantafyllidis-EtAl-Surface-instability-of-an-elastic-2008},
or rod-like solids with large incompatible strain~\cite{KB1,KB2,%
Lestringant-Audoly-Elastic-rods-with-2016}, with applications ranging
from morphogenesis~\cite{Savin-Kurpios-EtAl-On-the-growth-and-form-of-the-gut-2011} to the active control of surface
properties~\cite{Terwagne-Brojan-EtAl-Smart-morphable-surfaces-2014}.
An important exception to this rule is when the bifurcation problem
has no intrinsic length-scale as happens for a compressed
hyperelastic block, a problem considered by Biot~\cite{Biot63,biot}: a
continuum of linear modes appear simultaneously at the bifurcation
threshold with all possible wavelengths. This free-surface
instability has been characterized numerically and experimentally only
recently, and was found to be subcritical, localized and non-linear
in essence~\cite{HongSuo2009,Yoon_Kim_Hayward, Hohlfeld-Mahadevan-Unfolding-the-sulcus-2011,
Mora-Abkarian-EtAl-Surface-instability-of-soft-2011,%
Hohlfeld-Mahadevan-Scale-and-Nature-of-Sulcification-2012,%
Cai_Chen_2012,C3SM51512E,Jin_2015,Jin2015}.
In spite of recent
progress~\cite{Cao-Hutchinson-From-wrinkles-to-creases-2011,%
Fu-Ciarletta-Buckling-of-a-coated-elastic-2015}, there is no simple
and systematic theoretical argument that explains why and in which
circumstances localized creasing patterns are to be observed, nor whether
scale-invariance is a sufficient condition for localization.

Here, we analyze a variant of Biot's compressed elastic block, in
which we replace the half-space geometry by a prism.  The ridge angle
$\phi$ brings in an additional parameter.  Experimentally, we find
buckling patterns reminiscent of creasing when the prism is flat
enough ($\phi$ close to $180^\circ)$, consistent with prior work~\cite{HongSuo2009,Yoon_Kim_Hayward, Hohlfeld-Mahadevan-Unfolding-the-sulcus-2011,
Mora-Abkarian-EtAl-Surface-instability-of-soft-2011,%
Hohlfeld-Mahadevan-Scale-and-Nature-of-Sulcification-2012,%
Cai_Chen_2012,C3SM51512E,Jin_2015,Jin2015}.  For
acute enough ridge angles, however, a smooth buckling mode develops
near the ridge, with a well-defined wavelength.  We carry out a linear
stability analysis of the compressed hyperelastic prism and
investigate the competition between smooth and localizing buckling
modes.

\begin{figure}
\centering
\includegraphics[width=.99\columnwidth]{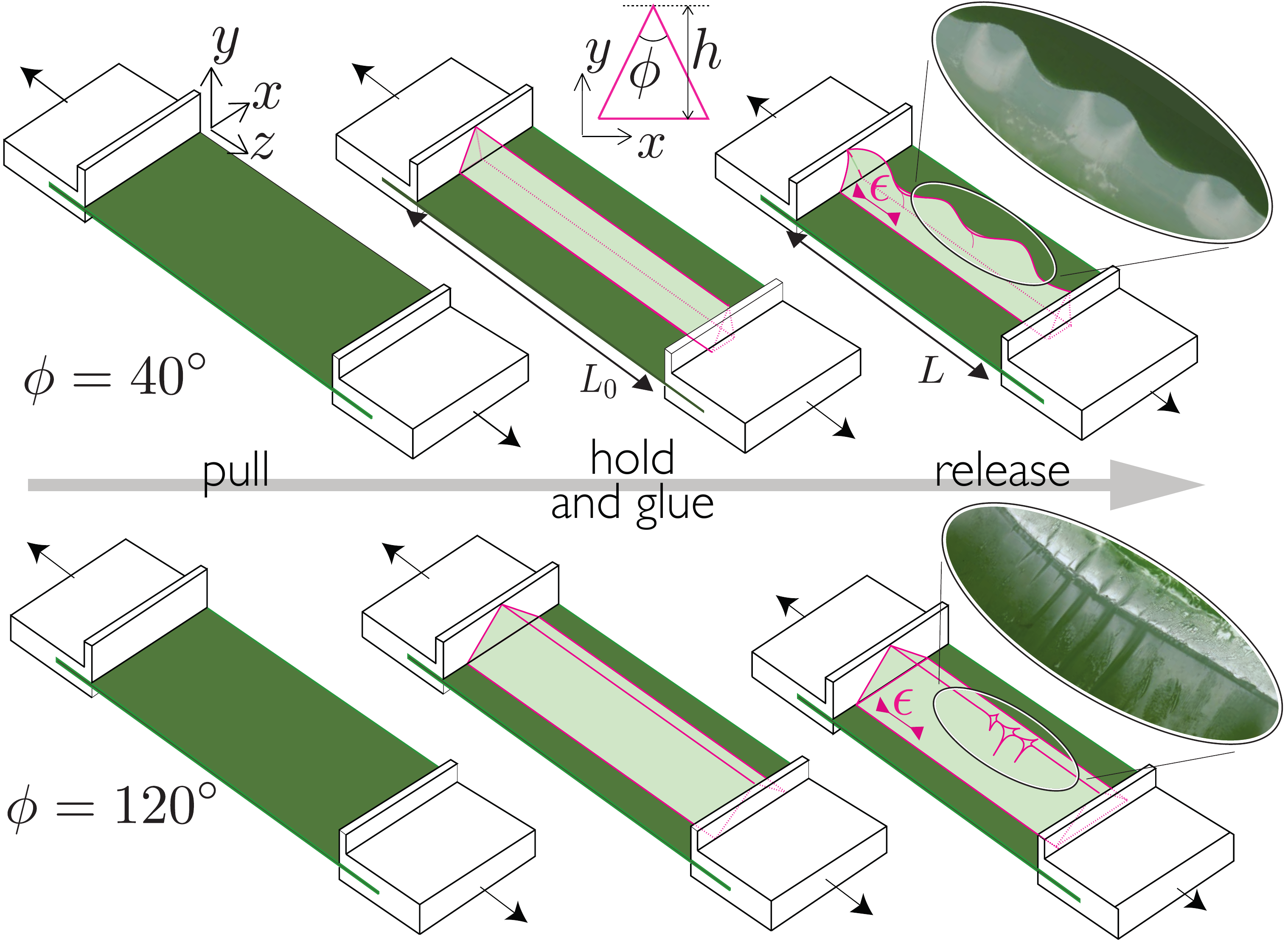}
\caption{Sketch of the experimental setup: to set the prism in axial
compression, we first strech the substrate, then glue the prism to the
substrate while keeping the substrate in tension, and finally release
the substrate.  This induces buckling of the prism: extended wrinkling
(top row, $\phi = 40^\circ$) and localized creasing (bottom row, $\phi
= 120^\circ$) are observed, depending on the value of $\phi$. Insets: 
experimental pictures.
\label{fig:intro}}
\end{figure}

In our experiments, we use an isosceles-triangular prism made of a
silicon elastomer (Ecoflex).  This elastomer is nearly incompressible
with a Young's modulus $E_\text{p} \approx 0.06 \pm 0.02 \text{MPa}$.
Its lower face is bonded to a parallepipedic silicone block made of
vinylpolysiloxane whose Young's modulus is $\sim 20$ times larger,
$E_\text{b} \approx 1.3 \pm 0.05 \text{MPa}$.  Both the prism and the
base are obtained by casting liquid polymer into molds made of PMMA
obtained by laser-cutting.  We stretch the base to a length $L_{0}$
prior to glueing the prism onto it, see Fig.~\ref{fig:intro}.  By
bringing the ends of the base closer to one another we induce a
compressive axial strain $\epsilon = \frac{L_0-L}{L_0}$ in the prism,
that depends on the current length $L<L_{0}$ of the base.  At a
critical value of the strain $\epsilon_{\mathrm{c}}$ an instability is
observed which is localized along the free ridge of the prism,
opposite to the base.  Using different molds we repeat the buckling
experiment for different ridge angles $\phi$ in the range $20^\circ$
to $120^\circ$.  The height $h$ of the triangular prism is chosen at
least ten times smaller than $L_{0}$, so we can ignore finite-length
effects in the analysis.

\begin{figure}
\centering
\includegraphics[width=.95\columnwidth]{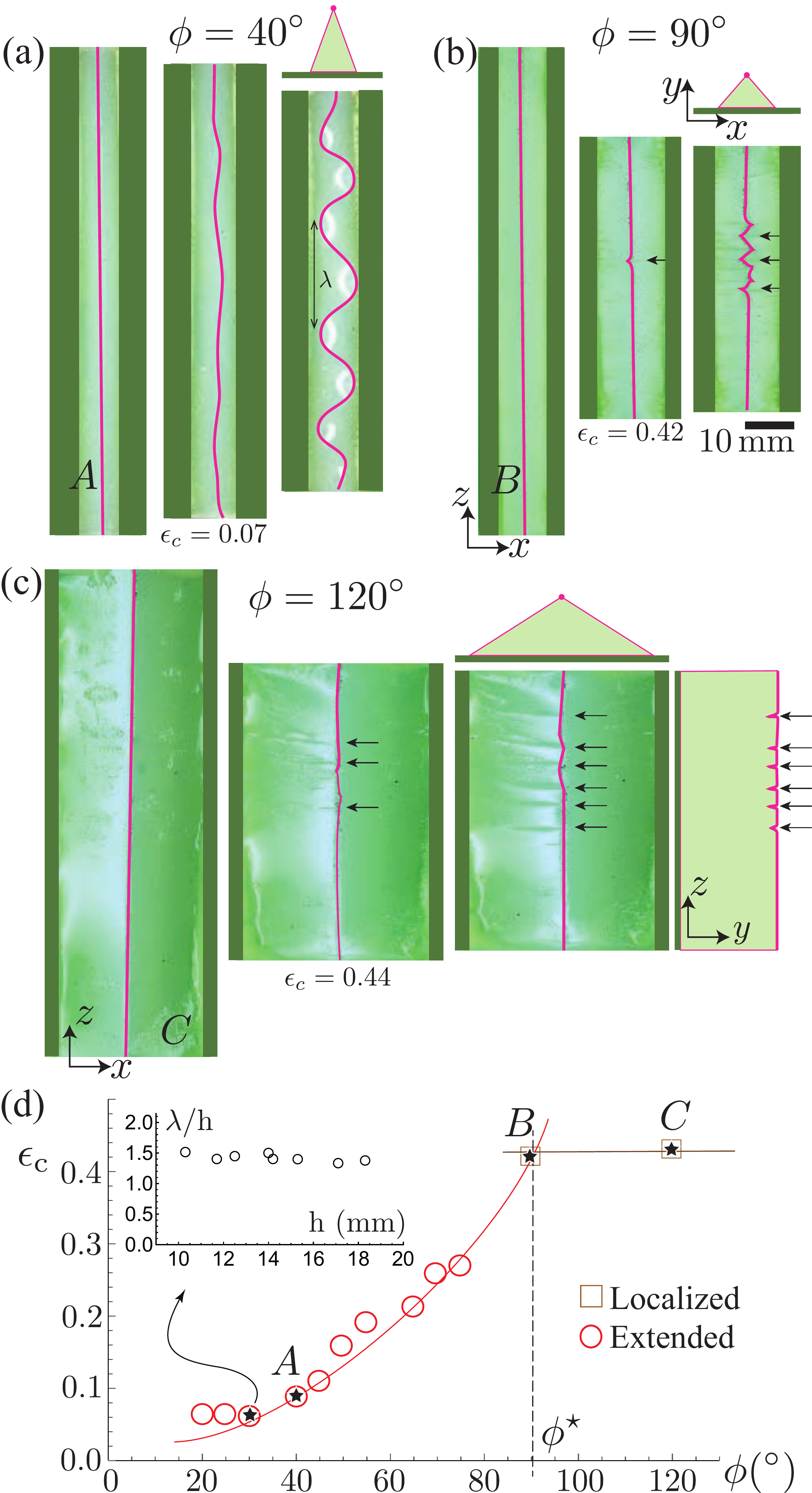}
\caption{Experimental results.  (a--c) Top-views in the $(x,z)$-plane.
Each set of pictures is for increasing axial strain $\epsilon$.
$\text{A}$: $\phi = 40^\circ$, $h = 10\, \mathrm{mm}$ and $L_0 = 100
\, \mathrm{mm}$, $\text{B}$: $\phi = 90^\circ$, $h = 5 \, \mathrm{mm}$
and $L_0 = 100 \, \mathrm{mm}$, and $\text{C}$: $\phi = 120^\circ$, $h
= 10 \, \mathrm{mm}$ and $L_0 = 100 \, \mathrm{mm}$.  Black arrows
highlight the creases visible on the faces.  (d) Critical strain
$\epsilon_c(\phi)$.  The thin, hand-drawn curves reveal the trend of
the experimental data points.  Inset: rescaled wavelength of the
extended mode $\lambda/h$
for $\phi = 30^{\circ}$.\label{fig:expresults}}
\end{figure}

For $\phi$ smaller than a critical value $\phi^\star \approx 90^\circ$
we observe a smooth, extended buckling mode whereby the ridge bends
out of the plane of symmetry of the prism, see
Fig.~\ref{fig:expresults}a;
we will refer to this as \textit{antisymmetric wrinkling} (AW).  For a
given angle $\phi\leq \phi^\star$ the wavelength $\lambda$ scales
close to linearly with the height of the prism $h$ for the range of
heights tested in the experiments, see the inset of
Fig.~\ref{fig:expresults}.

For $\phi\geq \phi^\star$ the buckling mode is entirely different, see
Fig.~\ref{fig:expresults}c: localized creases are initiated at the
ridge.  As the strain is increased beyond
$\epsilon_{\mathrm{c}}$, more creases are formed and they spread along
the lateral faces toward the base.  The gap between successive creases
does not appear to be regular.  This buckling mode will be referred to
as \textit{surface creasing} (SC).

Overall, $\epsilon_c$ increases steadily with the ridge angle $\phi$
until it reaches a plateau at $\epsilon_c \approx 0.42$ for $\phi =
\phi^\star$, where the nature of the buckling mode changes, see
Fig.~\ref{fig:expresults}d. This value is lower than the critical Biot strain
$\epsilon_{\mathrm{Biot}} = 0.55$ calculated by Biot~\cite{Biot63,biot} 
for the surface instability, as explained below.

We set up a bifurcation analysis, with the aim of characterizing the
instabilities and of explaining the competition between the localized
and extended buckling modes.  The system is modeled as an infinitely
long prism with triangular cross-section $\mathcal{D}$ made of an
hyperelastic material.  Its elastic energy density is denoted by
$W_{3D}(\bm{E})$ where $\bm{E}=\frac{1}{2}(\bm{F}^{T}\cdot \bm{F} -1)$
is the strain tensor, $\bm{F} = \frac{\partial (x, y, z)}{\partial
(X,Y,Z)}$ the transformation gradient, $(X,Y,Z)$ are the coordinates in
reference configuration with $Z$ aligned with the prism axis and
$Y$ along the axis of symmetry of the triangular cross-section
$\mathcal{D}$, and $(x,y,z)$ are the coordinates in deformed
configuration.  The expression of $W_{3D}(\bm{E})$ reflects the choice
of a material law; we use a Gent model, as described in the
supplementary material, with a choice of material parameters that
makes this constitutive law practically equivalent to an
incompressible neo-Hookean model.  Working in the framework of finite
elasticity, we denote by $\boldsymbol{\varphi}(X, Y, Z) = (x,y,z) -
(X,Y,Z) $ the displacement.  The non-linear equilibrium is obtained by
the principle of virtual work as
\begin{equation}
\forall \widehat{\boldsymbol{\varphi}}(X,Y,Z), \,
\int_{0}^{L_0}
\iint_{\mathcal{D}}
\bm{\Sigma}
:
\bigl(
\bm{F}^T\cdot
\widehat{\bm{F}}
\bigr)
\,\mathrm{d}X\,\mathrm{d}Y
\,\mathrm{d}Z
=
0
\textrm{,}
\label{eq:3DnonlinearEquil}
\end{equation}
where $\bm{\Sigma} = \frac{\partial W_{3D}}{\partial \bm{E}}$ denotes the
stress and $\widehat{\bm{F}} =\frac{\partial
\widehat{\boldsymbol{\varphi}}}{\partial (X,Y,Z)}$ the virtual
increment of deformation gradient.  As the average strain $\epsilon$ is
imposed by the base, we consider only admissible virtual displacements
$\widehat{\boldsymbol{\varphi}}$ whose incremental axial strain is
zero on average.  Taking advantage of the fact that the buckling
patterns are localized near the ridge in the experiments, we simplify
the boundary conditions at the interface with the base which we
replace by a free boundary.

The unbuckled solution is in a state of homogeneous `simple'
compression as described by $\boldsymbol{\varphi}_0^\epsilon = \eta(\epsilon)
\,\bigl(X \, \mathbf{e}_x + Y \, \mathbf{e}_y\bigr) -
\epsilon\,Z\,\mathbf{e}_{z}$. Here $\eta(\epsilon)$ captures the
dilation of the cross-section by Poisson's effect and is found from
the constitutive law by solving $\frac{\partial W_{3D}}{\partial
\eta}\bigl(\epsilon, \eta(\epsilon)\bigr)=0$.
We consider a small perturbation $\boldsymbol{\varphi}_{1}$, to this
invariant solution $\boldsymbol{\varphi}_{1}= \bigl( \xi_x(X,Y) \,
\mathbf{e}_x + \xi_y(X,Y) \, \mathbf{e}_y +
i\,\xi_z(X,Y)\,\mathbf{e}_{z} \bigr)\,e^{i\,q\,Z}$ in the form of a
pure Fourier mode with wavenumber $q$.  The virtual displacement
$\widehat{\boldsymbol{\varphi}}$ is sought in a similar form.  Upon
linearization and discretization using the finite-element method, the
equation of equilibrium~(\ref{eq:3DnonlinearEquil}) takes the form
\begin{equation}
\forall \widehat{\boldsymbol{\xi}},\quad
\widehat{\boldsymbol{\xi}}
\cdot
\bigl(
K_{\epsilon}
+
q\,C_{\epsilon}
+
q^2\,M_{\epsilon}
\bigr)
\cdot \boldsymbol{\xi}_1
= 0
\textrm{,}
\label{linear_stability}
\end{equation}
where $\boldsymbol{\xi}_{1} = \left( \xi_x, \, \xi_y, \, \xi_z\right)$
and $ \hat{\boldsymbol{\xi}} = \left( \widehat{\xi}_x, \,
\widehat{\xi}_y,\, \widehat{\xi}_z\right)$ are two vectors collecting
the Fourier amplitudes of the real and virtual nodal displacements on
the cross-section $\mathcal{D}$.  The Fourier analysis thus yields a
2-d eigenvalue problem in which the third dimension enters through the
wavenumber $q$ only.  For details of this 2-d formulation and of its
implementation, see Supplemental Material (Supp.\ Mat.)
and~\cite{Lestringant-Audoly-Elastic-rods-with-2016}.  To discretize
and solve the eigenvalue problem, we make use of the finite-element
library FEniCS~\cite{FEniCS} and of the SLEPc library~\cite{SLEPc}.

Equation~(\ref{linear_stability}) is invariant when a homothety is
applied to both the solution and the wavelength $2\pi/q$; in addition, the
domain $\mathcal{D}$ is scale-invariant near the tip (ridge).  As a
result, an infinite number of modes that are homothetic one to another
appear concurrently at the critical strain $\epsilon_{\mathrm{c}}$.
These modes are localized near the ridge and are associated with all
possible wavenumbers: there is no selection of the wavenumber in this
scale-invariant linear bifurcation analysis, see Supp.\ Mat.\ for
details.
By contrast, the critical strain $\epsilon_{\mathrm{c}}$ and the shape
of the buckling mode (up to a dilation) are selected as a function of
$\phi$.

\begin{figure}
\centering
\includegraphics[scale=0.25]{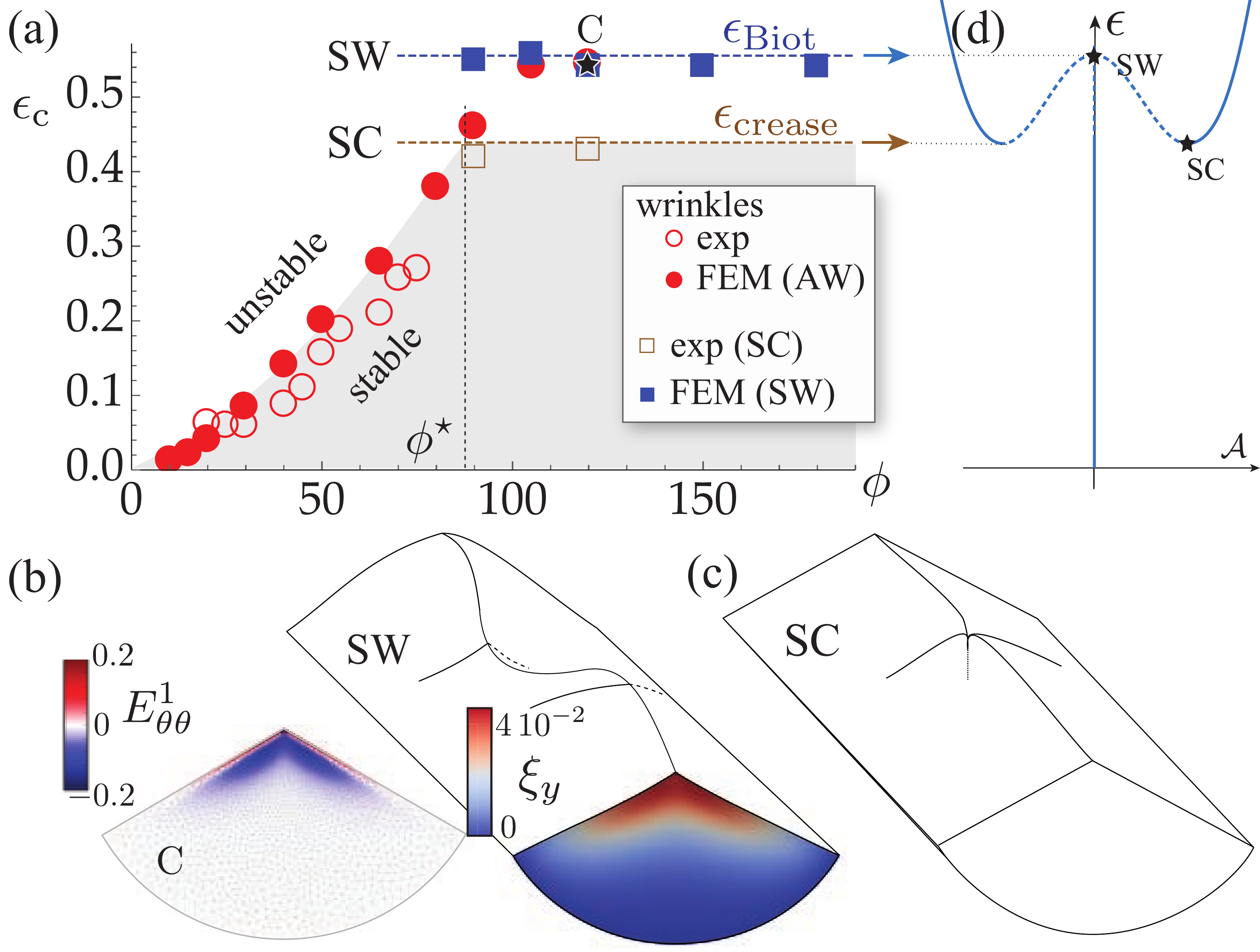}
\caption{ (a) Full bifurcation diagram, comparing the modes
predicted by the linear bifurcation analysis (AW and SW), Biot's
threshold, the non-linear creasing threshold, and experiments.
(b) Numerical linear buckling mode for $\phi = 120^{\circ}$, $h=
0.5$.  Sketch of the deformed prism superimposed with a colormap
of the amplitude of the vertical displacement $\xi_y$.  Colormap
of the amplitude of the hoop strain $E^1_{\theta\theta}$.  (c)
Sketch of the experimental surface creasing mode (SC) for $\phi =
120^{\circ}$.  (d) Sketch of the sub-critical bifurcation curve
$\mathcal{A}(\epsilon)$ for creasing.\label{fig:non_linear}}
\end{figure}

As the unbuckled configuration is mirror-symmetric with respect to the
$(yz)$ plane, the buckling modes can be either symmetric or
anti-symmetric.  When $\phi$ is smaller than $\approx 105^\circ$, the
first critical buckling mode predicted by the FEM analysis is an
anti-symmetric wrinkling mode (AW), see Fig.~\ref{fig:non_linear}a. It involves lateral undulations
of the ridge, see Fig.~\ref{fig:large_angles}b, similar to the
buckling mode seen in the experiments.  The corresponding critical strain
$\epsilon_{\mathrm{c}}$ is plotted in Fig.~\ref{fig:large_angles}a
(disks) and compared to experimental results (open circles):
\begin{figure}
\centering
\includegraphics[width=.99\columnwidth]{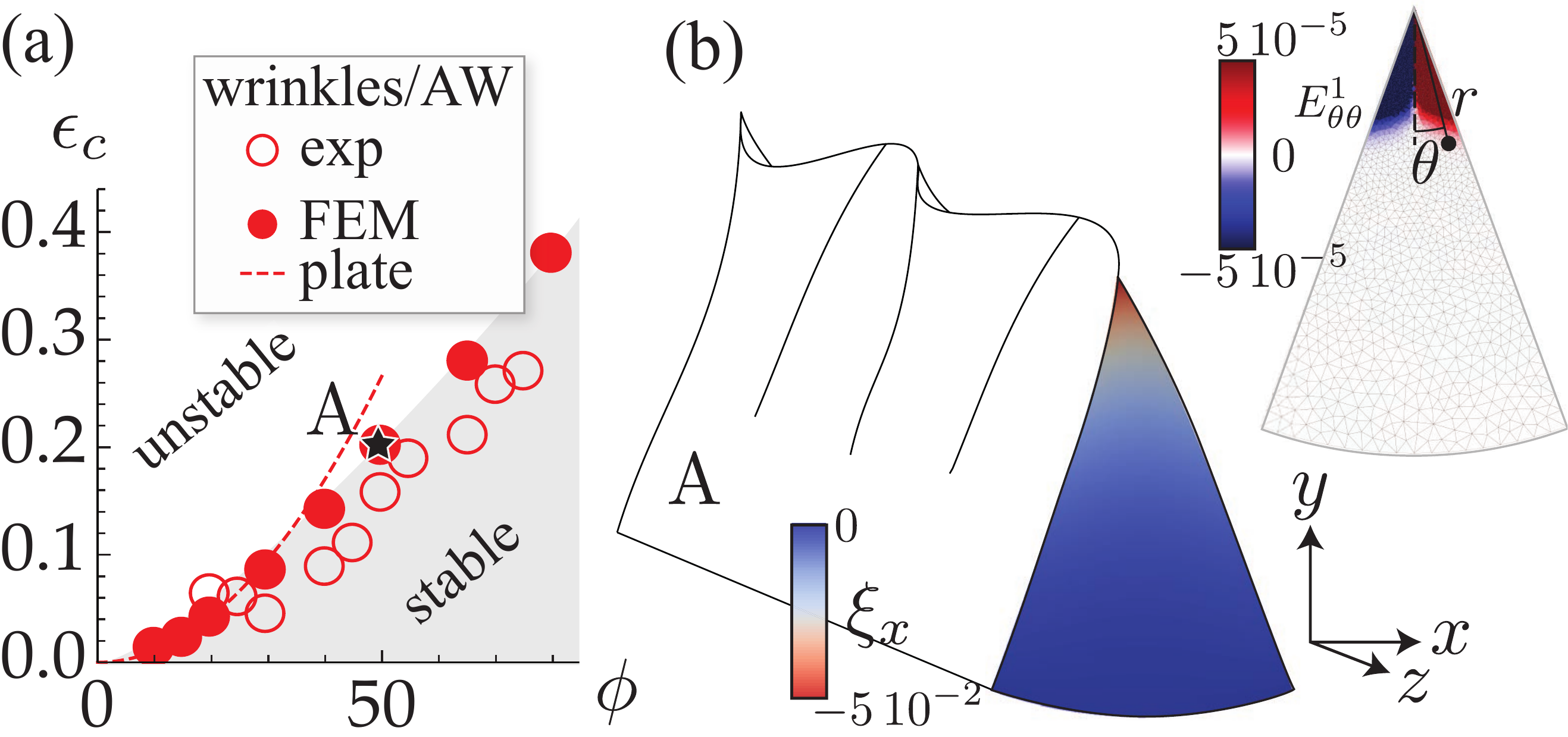}
\caption{Anti-symmetric wrinkling (AW).  (a) Phase diagram
$\epsilon_c(\phi)$ from experiments (open circles), simulations
(disks) and analytical model (dashed curve).  (b) Numerical
buckling mode for $\phi = 40^{\circ}$, $h= 5\, \mathrm{mm}$, shown with an
arbitrary amplitude.  The two colormaps show the amplitude of the
lateral displacement (left) and of the incremental hoop strain
$E^1_{\theta\theta}$ (right).\label{fig:large_angles}}
\end{figure}
$\epsilon_{\mathrm{c}}(\phi)$ is in good agreement with
the experiments, and increases with $\phi$.

In the limit of an acute ridge angle, $\phi\to 0$, the prism can be
modeled as a thin, infinitely long plate whose thickness $t(y)$ varies
linearly with the distance to the ridge, $t= \phi\, |h-y|$.  For the
unbuckled solution, the mid-surface of the plate is contained in the
$(yz)$ plane and has an axial pre-stress $\sigma^{0} = E \, \epsilon$.
When linearized about this solution, the F\"oppl-Von K\`arm\`an
equations for elastic plates yield, see for instance
\cite{Audoly-Pomeau-Elasticity-and-geometry:-from-2010},
\begin{equation}
(m_{\alpha \beta})_{,\alpha \beta} + t(y)\sigma^{0} w_{,zz}=0
\label{eq:FVK}
\end{equation}
where $w(x,y)$ is the (horizontal) deflection, $m_{\alpha \beta}= D(y)
\big((1-\nu)w_{,\alpha \, \beta} + \nu \delta_{\alpha \, \beta}
w_{,\gamma\, \gamma}\big)$ denotes the bending moment, $D (y)= \frac{E
\, t(y)^3}{12 (1-\nu^2)}$ is the bending modulus of the plate, $E$ is
Young's modulus and $\nu$ is Poisson's ratio.  A comma in subscript
denotes a partial derivative and Greek symbols are restricted to
in-plane directions, $\alpha, \beta \in \{y,z \}$.  We use Einstein's
convention for implicit summation on repeated indices.

In the plate model, we consider perturbations that are harmonic in the
axial direction and rescale the vertical coordinate using the
wavelength, $w(y) = \overline{w}(q \, y)\, e^{i \, q \, z}$.  When
expressed in terms of $\overline{w}$ and $q$, the boundary value
problem~(\ref{eq:FVK}) and the associated boundary conditions depend
on the two dimensionless parameters $\nu$ and $\overline{\sigma}^{0} =
\frac{12 \, (1-\nu^2) \, \sigma^0}{E\,\phi^2}$.  A numerical solution
based on a shooting method yields the critical value
$\overline{\sigma}^{0}_c(\nu)$, see Supp.\ Mat.\ for details.  The
corresponding critical strain is $\epsilon_{\mathrm{c}} = \frac{
\overline{\sigma}^{0}_c(\nu)}{12 \, (1-\nu^2)} \, \phi^{2}$.  For our
particular 3-d constitutive law, $\nu = 0.45$ and we obtain
$\epsilon_{\mathrm{c}} (0.45) \approx 0.35 \, \phi^2$; the dependence 
on Poisson's ratio is mild, $\epsilon_{\mathrm{c}} (0.50) \approx 
0.33 \, \phi^2$ in the incompressible case.  This
prediction has no adjustable parameter and is plotted in
Fig.~\ref{fig:large_angles}a (red dashed line): it agrees
asymptotically with the finite element analysis for $\phi\to 0$.  Note
that $\epsilon_c\sim \phi^2$ is small when $\phi \to 0$, which is
consistent with the linear elastic behavior assumed in the plate
model.

\emph{Symmetric wrinkling}  modes (SW) are also found in the numerical
bifurcation analysis.  They extend on the adjacent faces on both sides
of the ridge and involve an undulation of the ridge in the plane of
symmetry $(yz)$, see Fig.~\ref{fig:non_linear}b.  The strain at which
the first symmetric mode appears is $\epsilon \approx 0.55$, a value
which hardly depends on the ridge angle $\phi$, see
Fig.~\ref{fig:non_linear}a.
This value is consistent with the critical Biot strain
$\epsilon_{\mathrm{Biot}} = 0.55$ corresponding to the existence of a
marginally stable surface mode in a pre-stressed neo-Hookean
half-space~\cite{Biot63,biot}.  This is consistent with the fact that
the SW mode is localized just beneath the faces of prism, see
Fig.~\ref{fig:non_linear}b.  When $\phi$ reaches $\approx 105^\circ$,
the critical strain $\epsilon_{\mathrm{c}}$ of the AW mode becomes
larger than $\epsilon_{\mathrm{Biot}} = 0.55$: the numerical analysis
then predicts that the first buckling mode switches from an
antisymmetric mode (AW) to a symmetric (SW) mode, see
Fig.~\ref{fig:non_linear}a.

This \emph{linear} analysis therefore predicts a symmetric wrinkling mode (SW)
which is smooth and sinusoidal, in apparent contradiction with the
localized pattern observed in the experiments.  When this mode becomes
unstable, all wavelengths appear concurrently: it is known that the
non-linear coupling between the different wavelengths gives rise to a
creasing instability
through a sub-critical
bifurcation~\cite{Cao-Hutchinson-From-wrinkles-to-creases-2011,%
Hohlfeld-Mahadevan-Unfolding-the-sulcus-2011}.  The buckling strain
for the creasing instability in a neo-Hookean half-plane
$\epsilon_{\textrm{crease}}
\approx 0.44$ is therefore lower than that predicted by the linear
analysis $\epsilon_{\mathrm{Biot}} \approx 0.55$,
see~\cite{Hohlfeld-Mahadevan-Unfolding-the-sulcus-2011,%
Cai_Chen_2012}.  Extrapolating to our problem, this suggests that our
SW modes are subcritical as well, and that the critical strain
$\epsilon_{\mathrm{Biot}}$ predicted by the linear analysis needs to be corrected:
the value $\epsilon_{\textrm{crease}}$ has been included in
Fig.~\ref{fig:non_linear}a and indeed corresponds to the plateau
observed in the experiments, see Fig.~\ref{fig:non_linear}.
Accordingly, the critical ridge angle $\phi^\star$ can be found by
equating the critical strain for antisymmetric modes
$\epsilon_{\mathrm{c}}(\phi)$ with the creasing strain
$\epsilon_{\textrm{crease}}$:
this yields $\phi^\star = 88^\circ$, see Fig.~\ref{fig:non_linear},
which accurately matches the experimental value $\phi^\star\approx
90^\circ$.

Our linear stability analysis correctly captures the dependence of the
critical strain on the ridge angle, $\epsilon_\mathrm{c}(\phi)$, as
well as the shape of the antisymmetric mode.  In our scale-free
formulation, there is no selection of the wavelength.  To account for
the wavelength of the antisymmetric mode, one would need to consider
additional ingredients in the analysis, such as subtle nonlinear
effects and/or small-scale regularization.  By contrast with the
antisymmetric mode, the symmetric mode predicted by the linear
stability analysis is not observed, as it gives rise to creasing by a
subcritical bifurcation.  Combining our linear analysis with the
non-linear threshold for creasing, we have explained the critical
value of the ridge angle $\phi^\star\approx 90^\circ$ at which the
pattern changes.  Interestingly, close to $\phi^\star$, the system
displays a mix of the two behaviors: creases superimposed onto the
smooth antisymmetric mode are shown in Fig.~\ref{fig:expresults}b,
probably resulting from the non-linear interaction between the
symmetric and antisymmetric modes.

The creasing localization has been explained in earlier work by
non-linear coupling of the buckling modes.  A remarkable finding of
our experiments is that our system features both localized creases and
a smooth extended buckling pattern: the coupling between modes of different
wavelengths is effective for the symmetric mode (leading to creases)
but it is not effective for the antisymmetric mode, surprisingly.
Therefore, scale-invariance in not a sufficient condition for
localization and the exact conditions in which modes of different
wavelengths can cooperate remain to be elucidated: the compressed
hyperelastic prism provides a workbench for future non-linear analyses
of creasing.

Acknowledgments: We thank A.~El Ouardy for his contribution to the
experiments.  CM acknowledges the financial support of Project
ANR-13-JS09-0009 (Agence Nationale de la Recherche, 2014).




\narrowtext
\end{document}